\documentclass[final,twocolumn,showpacs,amsmath]{revtex4}
\usepackage{graphicx}
\usepackage[T1]{fontenc}

\begin{document}

\title{Theory of High-Field Transports in Metallic Single-Wall Nanotubes}

\author{S.~Fujita}
\email{fujita@buffalo.edu}
\affiliation{Department of Physics, SUNY at Buffalo, Buffalo, New York 14260, USA}

\author{H.\,C.~Ho}
\email{hcho@sincerelearning.hk}
\affiliation{Sincere Learning Centre, Kowloon, Hong Kong, China}

\begin{abstract}
Individual metallic single-wall carbon nanotubes show unsual non-Ohmic
transport behaviors at high bias fields. For low resistance contact
samples, the differential conductance $dI/dV$ increases with increasing
bias, reaching a maximum at $\sim 100$\,mV. As the bias increases
further, $dI/dV$ drops dramatically [Yao \emph{et~al}.,
  \emph{Phys.\ Rev.\ Lett.}\ \textbf{84}, 2941, 2000]. The higher the
bias, the system behaves in a more normal (Ohmic) manner. This so-called
zero-bias anomaly is temperature-dependent ($50$--$150$\,K). We propose a
new interpretation. Supercurrent runs in the graphene wall below $\sim
150$\,K. The normal conduction-electron currents run outside the wall,
which are subject to the scattering by phonons and impurities. The
currents along the tube induce circulating magnetic fields and
eventually destroy the supercurrent in the wall at high enough bias, and
restore the Ohmic behavior. If the prevalent ballistic electron model is
adopted, then the scattering effects cannot be discussed.
\end{abstract}


\maketitle

\section{INTRODUCTION}

\begin{figure}
\centering
\includegraphics{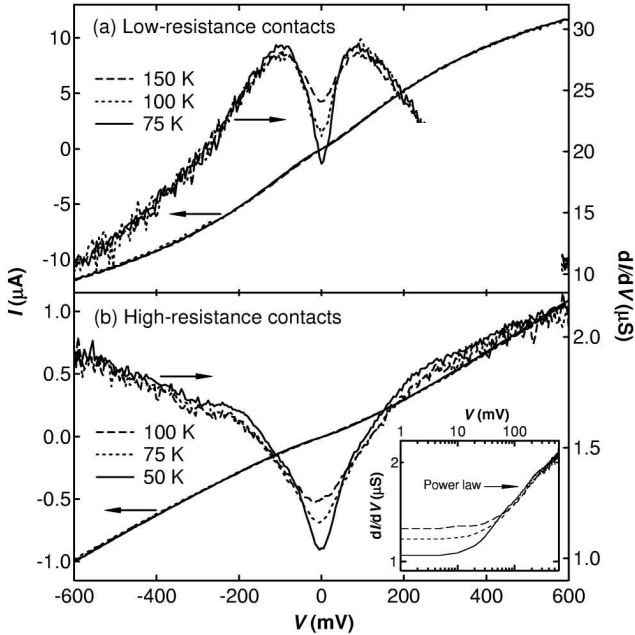}
\caption{Typical current $I$ and differential conductance $dI/dV$ vs
  voltage $V$ obtained using (a) low-resistance contacts (LRC) and (b)
  high-resistance contacts (HRC). The inset to (b) plots $dI/dV$ vs $V$
  on a double-log scale for the HRC sample. After Yao
  \emph{et~al}.\ \cite{Yao}.}\label{fg:Figure1}
\end{figure}

In 2000, Yao, Kane and Dekker \cite{Yao} reported the high-field transports in
metallic Single-Wall carbon NanoTubes (SWNT). In Fig.~\ref{fg:Figure1}, we reproduced
their $I$-$V$ curves, after Ref.~\cite{Yao}, Fig.~1.%
At low fields (voltage $\sim 30$\,mV), the currents show
temperature-dependent dips near the origin, exhibiting non-Ohmic
behaviors while at the high fields ($\sim 5$\,V) the resistance $R$
versus the bias voltage $V$ shows a relation:
\begin{equation}
R = R_0 + V/I_0\quad (\text{high } V),
\label{eq:vsa_relation}
\end{equation}
where $R_0$ and $I_0$ are constants. The original authors discussed the
low-field behavior in terms of one-dimensional (1D) Luttinger liquid
(LL) model. Many experiments however indicate that the electrical
transports in SWNT have a two-dimensional (2D) character.~\cite{Saito} In
fact, the conductivity in individual nanotubes depends on the
circumference and the pitch characterizing a space-curve (2D). Hence the
nanotube physics requires a 2D theory. In the present work, we present a
unified microscopic theory of both low- and high-field conductivities.
Carbon nanotubes are discovered by Iijima.~\cite{Iijima1} The important
questions are how the electrons or other charged particles traverse the
nanotubes and whether these particles are scattered by impurities and
phonons or not. To answer these questions, we need the electron energy
band structures. Wigner and Seitz (WS) \cite{Wigner} developed the WS
cell model to study the ground state of a metal. Starting with a given
lattice, they obtain a Brillouin zone in the $k$-space and construct a
Fermi surface. This method has been successful for cubic crystals
including the face-centered cubic (fcc), the body-centered cubic (bcc),
diamond and zincblende lattices. If we apply the WS cell model to
graphene, we then obtain a gapless semiconductor, which is not
experimentally observed.\ \cite{Saito} We will overcome this difficulty
by taking a different route in Sec.~\ref{se:Graphene}.

SWNTs can be produced by rolling graphene sheets into circular cylinders
of about one nanometer (nm) in diameter and microns ($\mu$m) in
length.\ \cite{Iijima2} The electrical conduction in SWNTs depends on
the circumference and pitch, and can be classified into two groups:
either semiconducting or metallic.\ \cite{Saito} In our previous work
\cite{Fujita1}, we have shown that this division in two groups arises as
follows. A SWNT is likely to have an integral number of carbon hexagons
around the circumference. If each pitch contains an integral number of
hexagons, then the system is periodic along the tube axis, and ``holes''
(not ``electrons'') can move along the tube. Such a system is
semiconducting and its electrical conductivity increases with the
temperature, and is characterized by an activation energy
$\varepsilon_3$.\ \cite{Fujita2} The energy $\varepsilon_3$ has a
distribution since both pitch and circumference have distributions. The
pitch angle is not controlled in the fabrication processes. There are
far more numerous cases where the pitch contains an irrational number of
hexagons. In these cases, the system shows a metallic behavior
experimentally observed.\ \cite{Moriyama}

\begin{figure}
\centering
\includegraphics{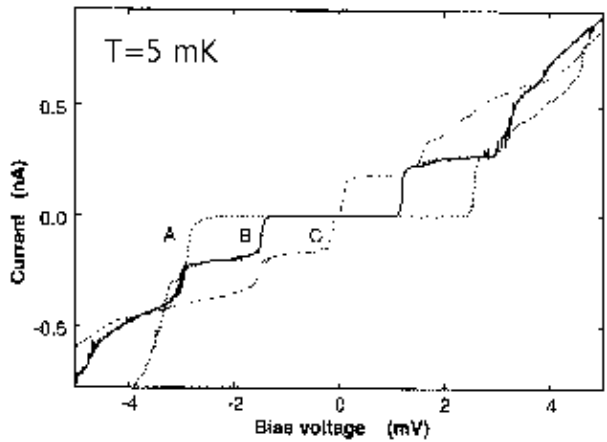}
\caption{Current-voltage curves of metallic SWNT at gate voltages of $88.2$\,mV (trace A), $104.1$\,mV
(trace B) and $120.0$\,mV (trace C). After Tans \emph{et al}.\ in Ref.~\cite{Tans1}, Fig.~2.}\label{fg:Figure2}
\end{figure}

We primarily deal with the metallic SWNTs in the present work. Before
dealing with high-field transports, we briefly discuss the low-field
transports. Tans \emph{et al}.\ \cite{Tans1} measured the electrical
currents in metallic SWNTs under bias and gate voltages. Their data from
Ref.~\cite{Tans1}, Fig.~2, are reproduced in
Fig.~\ref{fg:Figure2}.%
The currents versus the bias voltage are plotted in Fig.~\ref{fg:Figure2} at three gate voltages:
A ($88.2$\,\,mV), B ($104.1$\,mV), C ($120.0$\,mV). Significant features are:
\begin{enumerate}
\item[(i)] A non-Ohmic behavior is observed for all, that is, the
  currents are not proportional to the bias voltage except for high
  bias. The gate voltage charges the tube. The Coulomb (charging) energy
  of the system having charge $Q$ is represented by
\begin{equation}
E_\text{Coul} = Q^2/2C,
\label{eq:Qir_by}
\end{equation}
where $C$ is the total capacitance of the tube.

\item[(ii)] The current near the origin appears to be constant for
  different gate voltages $V_\text{gate}$, (A)--(C). This feature was
  confirmed by later experiments.\ \cite{Tans2,Moriyama} The current
  does not change for small varying gate voltage in a metallic SWNT
  (while the current (magnitude) decreases in a semiconducting SWNT).

\item[(iii)] The current at gate voltage $V_\text{gate} = 88.2$\,mV (A)
  reverts to the normal resistive behavior after passing the critical
  bias voltages on both (positive and negative) sides. Similar behaviors
  are observed for $V_\text{gate} = 104.1$\,mV (B) and $V_\text{gate} =
  120.0$\,mV (C).

\item[(iv)] The flat current is destroyed for higher bias voltages
  (magnitudes). The critical bias voltage becomes smaller for higher
  gate voltages.

\item[(v)] There is a restricted $V_\text{gate}$-range (view window) in
  which the horizontal stretch can be observed.
\end{enumerate}
Tan \emph{et al}.~\cite{Tans1} interpreted the flat currents near
$V_\text{bias} = 0$ in Fig.~\ref{fg:Figure2} in terms of a ballistic
electron model \cite{Saito}.

We propose a differing interpretation of the data in
Fig.~\ref{fg:Figure2} based on the Cooper pair \cite{Cooper} (pairon)
carrier model. Pairons move as bosons, and hence they are produced with
no activation energy factor. All features (i)--(v) can be explained
simply with the assumption that the nanotube wall is in the
\emph{superconducting state} as explained below.

The supercurrent runs without obeying Ohm's law. This explains the
feature (i). The supercurrents can run with no resistance due to the
phonon and impurity scattering and with no bias voltage. Bachtold
\emph{et~al}.\ \cite{Bachtold} observed by scanned probe microscopy that the
currents run with no voltage change along the tube in metallic SWNTs.
The system is in a superconducting ground state, whose energy $-E_g$ is
negative relative to the ground-state energy of the Fermi liquid
(electron) state. If the total energy $E$ of the system is less than the
condensation energy $E_g$:
\begin{equation}
E = K + E_\text{Coul} + E_\phi \leq E_g ,
\label{eq:total_energy}
\end{equation}
where $K$ is the kinetic energy of the conduction electrons and the pairons, and
\begin{equation}
E_\phi = QV_\text{bias}
\end{equation}
is the Coulomb field energy, then the system is stable. Experiments in
Fig.~\ref{fg:Figure2} were done at $5$\,mK. Hence, we may drop the kinetic
energy $K$ hereafter. The superconducting state is maintained and the
currents run unchanged if the bias voltage $V_\text{bias}$ is not too
large so that the inequality \eqref{eq:total_energy} holds. This
explains the horizontal stretch feature (ii).

If the bias voltage is high enough so that the inequality symbol in Eq.~\eqref{eq:total_energy}
is reversed, then normal currents revert and exhibit the Ohmic behavior, which
explains the feature (iii).

The feature (iv) can be explained as follows. For higher $V_\text{gate}$ there are more
amounts of charge, and hence the charges $Q_\text{A}$, $Q_\text{B}$,
$Q_\text{C}$ for the three cases (A, B, C) satisfy the inequalities:
\begin{equation}
Q_\text{A} < Q_\text{B} < Q_\text{C}.
\label{eq:Q_inequalities}
\end{equation}
The horizontal stretches are longer for smaller bias voltages. At the end of the
stretch ($V_\text{bias,max}$) the system energy equals the condensation
energy $E_g$. Hence, we obtain from Eq.~\eqref{eq:total_energy} after
dropping the kinetic energy $K$
\begin{align}
E_{\phi,\text{max}} &= QV_\text{bias,max} \equiv QV_\text{max} \nonumber\\
&= E_g - E_\text{Coul} = E_g - Q^2 / 2C.
\end{align}
Using Eq.~\eqref{eq:Q_inequalities}, we then obtain
\begin{equation}
V_\text{max,A} > V_\text{max,B} > V_\text{max,C},
\end{equation}
which explains the feature (iv).

The horizontal stretch becomes shorter as the gate voltage $V_\text{gate}$ is raised; it
vanishes when $V_\text{gate}$ is a little over $120.0$\,mV. The limit is given by
\begin{equation}
E_{\phi,\text{max}} = E_g - E_\text{Coul} = E_g - Q^2 / 2C = 0.
\end{equation}
If the charging energy $E_\text{Coul}$ exceeds the condensation energy $E_g$, then there are
no more supercurrents, which explains the feature (v). Clearly the important physical property in our pairon model is the
condensation energy $E_g$.

In the currently prevailing theory~\cite{Saito}, it is argued that the
electron (fermion) motion becomes ballistic at a certain quantum
condition. But all fermions are known to be subject to scattering. It is
difficult to justify the reason why the ballistic electron is not
scattered by impurities and phonons, which naturally exist in nanotubes.
Yao, Kane and Dekker \cite{Yao} emphasized the importance of phonon
scattering effects in their analysis of their data in Fig.~\ref{fg:Figure1}.
The Cooper pairs \cite{Cooper} in supercurrents, as is known, can run with no
resistance (due to impurities and phonons). Clearly the experiments on
the currents shown in Fig.~\ref{fg:Figure1} are temperature-dependent,
indicating the importance of the electron-phonon scattering effect. If
the ballistic electron model is adopted, then the phonon scattering
cannot be discussed within the model's framework. We must go beyond the
ballistic electron model.

If the SWNT is unrolled, then we have a graphene sheet, which can be
superconducting at a finite temperature. We first study the conduction
behavior of graphene in Sec.~\ref{se:Graphene}, starting with the
honeycomb lattice and introducing ``electrons'' and ``holes'' based on
the orthogonal unit cell. Phonons are generated based on the same
orthogonal unit cell. In Sec.~\ref{se:Phonons}, we treat phonons and
phonon-exchange attraction. In Sec.~4, we construct a Hamiltonian
suitable for the formation of the Cooper pairs. We derive, in Sec.~5,
the linear dispersion relation for the center-of-mass motion of the
pairons. The pairons moving with a linear dispersion relation undergoes
a Bose-Einstein condensation (BEC) in 2D, which is shown in Sec.~6.
Zero-bias anomaly is discussed is Sec.~7.

\section{GRAPHENE}\label{se:Graphene}

Following Ashcroft and Mermin \cite{Ashcroft}, we adopt the
semiclassical model of electron dynamics in solids. In the semiclassical
(wave packet) theory, it is necessary to introduce a $k$-vector
\begin{equation}
\mathbf{k} = k_x\mathbf{\hat e}_x + k_y\mathbf{\hat e}_y + k_z\mathbf{\hat e}_z,
\end{equation}
$\mathbf{\hat e}_x$, $\mathbf{\hat e}_y$ and $\mathbf{\hat e}_z$ are Cartesian orthonormal vectors
since the $k$-vectors are involved in the semiclassical equation of
motion:
\begin{equation}
\hbar \dot{\mathbf{k}} \equiv \hbar \frac{d\mathbf{k}}{dt} = q(\mathbf{E} + \mathbf{v} \times \mathbf{B}),
\end{equation}
where $\mathbf{E}$ and $\mathbf{B}$ are the electric and magnetic fields, respectively,
and the vector
\begin{equation}
\mathbf{v} = \frac{\partial \varepsilon}{\partial \mathbf{k}}
\end{equation}
is the electron velocity where $\varepsilon$ is the energy. The
2D crystals such as graphene can also be treated similarly, only the
$z$-components being dropped. The choice of the Cartesian axes and the
unit cell is obvious for the cubic crystals. We must choose an orthogonal
unit cell also for the honeycomb lattice, as shown below.

\begin{figure}
\centering
\includegraphics{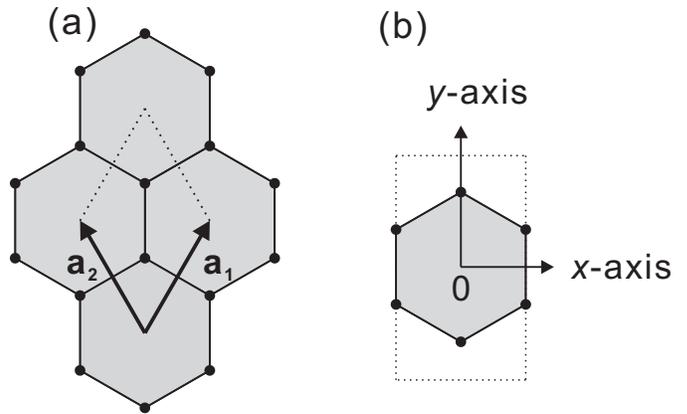}
\caption{(a) WS unit cell, rhombus (dotted lines) for graphene. (b) The
  orthogonal unit cell, rectangle (dotted lines).}
\label{fg:Figure3}
\end{figure}

Graphene forms a 2D honeycomb lattice. The WS unit cell is a rhombus
shown in Fig.~\ref{fg:Figure3}(a).%
The potential energy $V(\mathbf{r})$ is \emph{lattice-periodic}:
\begin{equation}
V(\mathbf{r} + \mathbf{R}_{mn}) = V(\mathbf{r}),
\label{eq:periodic_PE}
\end{equation}
where 
\begin{equation}
\mathbf{R}_{mn} \equiv m\mathbf{a_1} + n\mathbf{a_2}
\end{equation}
are Bravais vectors with the primitive vectors
($\mathbf{a_1},\mathbf{a_2}$) and integers ($m,n$). In the field
theoretical formulation, the field point $\mathbf{r}$ is given by
\begin{equation}
\mathbf{r} = \mathbf{r}\boldsymbol{'} + \mathbf{R_{nm}},
\end{equation}
where $\mathbf{r}\boldsymbol{'}$ is the point defined within the
standard unit cell. Equation \eqref{eq:periodic_PE} describes the 2D
lattice periodicity but does \emph{not} establish the $k$-space, which
is explained below.

To see this clearly, we first consider an electron in a simple square
(sq) lattice. The Schr\"odinger wave equation is
\begin{equation}
\mathrm{i}\hbar\frac{\partial}{\partial t}\psi(\mathbf{r}) =
-\frac{\hbar^2}{2m^*}\nabla^2\psi(\mathbf{r}) + V_\text{sq}(\mathbf{r})\psi(\mathbf{r}).
\label{eq:Schrodinger_equation}
\end{equation}
The Bravais vector for the sq lattice, $\mathbf{R}_{mn}^{(0)}$, is
\begin{align}
\mathbf{R}_{mn}^{(0)} &\equiv m\mathbf{a}_x + n\mathbf{a}_y \nonumber\\
&= ma\mathbf{\hat e}_x
+ na\mathbf{\hat e}_y,\quad(a = \text{lattice constant}).
\label{eq:Bravais_vector}
\end{align}
The system is lattice periodic:
\begin{equation}
V_\text{sq}\left(\mathbf{r} + \mathbf{R}_{mn}^{(0)}\right) = V_\text{sq}(\mathbf{r}).
\label{eq:lattice_periodic}
\end{equation}
If we choose a set of Cartesian coordinates $(x,y)$ along the sq
lattice, then the Laplacian term in Eq.~\eqref{eq:Schrodinger_equation} is given by
\begin{equation}
\nabla^2 \psi(x,y) = \left(\frac{\partial^2}{\partial x^2} + \frac{\partial^2}{\partial y^2} \right)
\psi(x,y).
\end{equation}
If we choose a periodic square boundary with the side length $Na$, $N=$
integer, then there are 2D Fourier transforms and (2D) $k$-vectors.

We now go back to the original graphene system. If we choose the
$x$-axis along either $\mathbf{a_1}$ or $\mathbf{a_2}$, then the
potential energy field $V(\mathbf{r})$ is periodic in the $x$-direction,
but it is aperiodic in the $y$-direction. For an infinite lattice the
periodic boundary is the only acceptable boundary condition for the
Fourier transformation. Then, there is no 2D $k$-space spanned by 2D
$k$-vectors. If we omit the kinetic energy term, then we can still use
Eq.~\eqref{eq:periodic_PE} and obtain the ground state energy (except
the zero point energy).

We now choose the orthogonal unit cell shown in Fig.~\ref{fg:Figure3}(b). The unit
has side lengths
\begin{equation}
b_1 = \sqrt{3} a_0,\qquad b_2 = 3a_0,
\end{equation}
where $a_0$ is the nearest neighbor distance between two C's. The unit
cell has 4 C's. The system is lattice-periodic in the $x$- and
$y$-directions, and hence there are 2D $k$-space.

The ``electron'' (``hole'') is defined as a \emph{quasi-electron} that
has an energy higher (lower) than the Fermi energy $\epsilon_\text{F}$
\emph{and} ``electrons'' (``holes'') are excited on the positive
(negative) side of the Fermi surface with the convention that the
positive normal vector at the surface points in the energy-increasing
direction.

The ``electron'' (wave packet) may move up or down along the $y$-axis to
the neighboring hexagon sites passing over one C$^+$. The positively
charged C$^+$ acts as a welcoming (favorable) potential valley for the
negatively charged ``electron'', while the same C$^+$ acts as a
hindering potential hill for the positively charged ``hole''. The
``hole'', however, can move horizontally along the $x$-axis without
meeting the hindering potential hills. Thus the easy channel directions
for the ``electrons'' (``holes'') are along the $y$-($x$-)axes.

Let us consider the system (graphene) at 0\,K. If we put an electron in
the crystal, then the electron should occupy the center O of the
Brillouin zone, where the lowest energy lies. Additional electrons
occupy points neighboring the center O in consideration of Pauli's exclusion
principle. The electron distribution is \emph{lattice-periodic} over the entire
crystal in accordance with the Bloch theorem.~\cite{Ashcroft}

Carbon (C) is a quadrivalent metal. The first few low-lying energy bands
are completely filled. The uppermost partially filled bands are
important for the transport properties discussion. We consider such a
band. The Fermi surface, which defines the boundary between the filled
and unfilled $k$-spaces (area) is not a circle since the $x$-$y$
symmetry is broken. The ``electron'' effective mass is lighter in the
$y$-direction than perpendicular to it. Hence the electron
motion is intrinsically anisotropic. The negatively charged ``electron''
is near the positive ions C$^+$ and the ``hole'' is farther away from
C$^+$. Hence, the gain in the Coulomb interaction is greater for the
``electron''. That is, the ``electron'' is more easily activated. Thus,
the ``electrons'' are the majority carriers at zero gate voltage.

We may represent the activation energy difference by~\cite{Fujita1}
\begin{equation}
\varepsilon_1 < \varepsilon_2.
\label{eq:activation-energy-difference}
\end{equation}
The thermally-activated (or excited) electron densities are given by
\begin{equation}
n_j(T) = n_j \mathrm{e}^{-\varepsilon_j / k_\text{B} T},
\label{eq:dag_by}
\end{equation}
where $j = 1$ and $2$ denote the ``electron'' and ``hole'',
respectively. The prefactor $n_j$ is the density at the high-temperature
limit.

\section{PHONONS AND PHONON EXCHANGE ATTRACTION}\label{se:Phonons}

Phonons are bosons corresponding to the running normal modes of the
lattice vibrations. They are characterized by the energy $\hbar\omega$,
where $\omega$ is the angular frequency, \emph{and} the momentum vector
$\hbar\mathbf{q}$, whose magnitude is $2\pi\hbar$ times the wave
numbers. The $q$-vector for phonons is similar to the $k$-vector for the
conduction electrons. The phonon with $\mathbf{q}$ represents a plane-wave
proceeding in the $\mathbf{q}$-direction. The frequency $\omega$ is
connected with the $q$-vector through the dispersion relation
\begin{equation}
\omega = \omega(\mathbf{q}).
\label{eq:dispersion_relation}
\end{equation}

The excitation of the phonons can be discussed based on the same
rectangular unit cell introduced for the conduction electrons. We note
that phonons can be discussed naturally based on the orthogonal unit
cells. [It is difficult to describe phonons in the WS cell model.] For
example, longitudinal (transverse) phonons proceeding upwards are
generated by imagining a set of plates each containing a number of
rectangular cells executing small oscillations vertically
(horizontally). A longitudinal wave proceeding in the crystal axis $x$,
is represented by
\begin{equation}
u_q \exp(-\mathrm{i}\omega_q t + \mathrm{i}\mathbf{q}\cdot\mathbf{r}) = u_q \exp(-\mathrm{i}\omega_q t + \mathrm{i}qx),
\label{eq:longitudinal_wave}
\end{equation}
where $u_q$ is the displacement in the $x$-direction. If we imagine a
set of parallel plates containing a great number of ions fixed in each
plate, then we have a realistic picture of the lattice vibration mode.
The density of ions changes in the $x$-direction. Hence, the
longitudinal modes are also called the \emph{density-wave} modes. The
transverse wave mode can also be pictured by imagining a set of parallel
plates containing a great number of ions fixed in each plate executing
the transverse displacements. Notice that this mode generates no
charge-density variation.

The Fermi velocity $v_\text{F}$ in a typical metal is of the order
$10^6$\,ms$^{-1}$ while the speed of sound is of the order
$10^3$\,ms$^{-1}$. The electrons are likely to move quickly to negate
any electric field generated by the density variations associated with
the lattice wave. In other words, the electrons may follow the lattice
waves instantly. Given a traveling normal wave mode in Eq.~\eqref{eq:dag_by}, we
may assume an electron density variation of the form:
\begin{equation}
C_\mathbf{q} \exp(-\mathrm{i}\omega_q t + \mathrm{i}\mathbf{q}\cdot\mathbf{r}).
\label{eq:eden_variation}
\end{equation}

Since electrons follow phonons immediately for all $\omega_q$,
the coefficient $C_\mathbf{q}$ can be regarded as independent of
$\omega_q$. If we further assume that the deviation is linear in the
scalar product $\mathbf{u}_q\cdot\mathbf{q}=qu_\mathbf{q}$ and again in
the electron density $n(\mathbf{r})$, we then obtain
\begin{equation}
C_\mathbf{q} = A_\mathbf{q}q u_\mathbf{q}n(\mathbf{r}).
\end{equation}
This is called the \emph{deformation potential approximation}.\ \cite{Harrison} The
dynamic response factor $A_\mathbf{q}$ is necessarily complex since the
traveling wave is represented by the exponential form. Complex
conjugation of Eq.~\eqref{eq:eden_variation} yields
$C_\mathbf{q}^*\exp(\mathrm{i}\omega_q t -
\mathrm{i}\mathbf{q}\cdot\mathbf{r})$. Using this form we can
reformulate the electron's response, but the physics must be the same.
From this consideration, we obtain
\begin{equation}
A_\mathbf{q} = A_{-\mathbf{q}}^*.
\end{equation}

Each normal mode corresponds to a harmonic oscillator characterized by
$(\mathbf{q}, \omega_q)$. The displacements $u_\mathbf{q}$ can be
expressed as
\begin{equation}
u_\mathbf{q} = \mathrm{i}\left(\frac{\hbar}{2\omega_q}\right)^{1/2}\left(a_\mathbf{q}-a_\mathbf{q}^\dagger\right),
\end{equation}
where $(a_\mathbf{q}, a_\mathbf{q}^\dagger)$ are operators satisfying
the \emph{Bose commutation rules}:
\begin{align}
\left[a_\mathbf{q}, a_\mathbf{p}^\dagger\right] &\equiv a_\mathbf{q}a_\mathbf{p}^\dagger - a_\mathbf{p}^\dagger a_\mathbf{q} = \delta_\mathbf{pq}, \nonumber\\
[a_\mathbf{q}, a_\mathbf{p}] &= \left[a_\mathbf{q}^\dagger, a_\mathbf{p}^\dagger\right] = 0.
\end{align}

Let us now construct an interaction Hamiltonian $H_\text{F}$, which has
the dimensions of an energy \emph{and} which is Hermitian. Using Eqs.
\eqref{eq:dispersion_relation} and \eqref{eq:longitudinal_wave}, we
obtain
\begin{equation}
H_\text{F} = \int d^3r\sum_\mathbf{q}\frac{1}{2}[A_\mathbf{q}q
  u_\mathbf{q}\exp(\mathrm{i}\mathbf{q}\cdot\mathbf{r})n(\mathbf{r}) + h.c],
\end{equation}
where \emph{h.c.} denotes the \emph{Hermitian conjugate}. This
Hamiltonian $H_\text{F}$ can be expressed as
\begin{align}
H_\text{F} &= \sum_\mathbf{k}\sum_\mathbf{q}\frac{1}{2}\left(V_q
c_{\mathbf{k}+\mathbf{q}}^\dagger c_\mathbf{k}a_\mathbf{q} + h.c.\right), \nonumber\\
V_q &\equiv A_q (\hbar / 2\omega_q)^{1/2}\mathrm{i}q,
\label{eq:interaction_Hamiltonian}
\end{align}
where $c$, $c^\dagger$ are electron operators satisfying the \emph{Fermi
  anticommutation rules}:
\begin{align}
\left\{c_\mathbf{k}, c_{\mathbf{k}\boldsymbol{'}}^\dagger\right\} &\equiv
c_\mathbf{k}c_{\mathbf{k}\boldsymbol{'}}^\dagger + c_{\mathbf{k}\boldsymbol{'}}^\dagger
c_\mathbf{k} = \delta_{\mathbf{k},\mathbf{k}\boldsymbol{'}}^{(3)}, \nonumber\\
\{c_\mathbf{k}, c_{\mathbf{k}\boldsymbol{'}}\} &= \left\{c_\mathbf{k}^\dagger,
c_{\mathbf{k}\boldsymbol{'}}^\dagger\right\} = 0.
\end{align}
The $H_\text{F}$ in Eq.~\eqref{eq:interaction_Hamiltonian} is the
\emph{Fr\"ohlich Hamiltonian}.\ \cite{Frohlich} In the process of
deriving Eq.~\eqref{eq:interaction_Hamiltonian}, we found that the
$H_\text{F}$ is applicable for the longitudinal phonons only. As noted
earlier, the transverse lattice normal modes generate no charge density
variations, making its contribution to $H_\text{F}$ negligible.

\section{THE FULL HAMILTONIAN}\label{se:Full_Hamiltonian}

Bardeen, Cooper and Schrieffer (BCS) published a historic theory of
superconductivity in 1957.\ \cite{Bardeen} Following BCS, Fujita and his
collaborators developed a quantum statistical theory of superconductivity in a
series of papers.\ \cite{Fujita5,Fujita6,Fujita7,Fujita8,Fujita9}
Following this theory, we construct a generalized BCS Hamiltonian in
this section.

In the ground state there are no currents for any system. To describe a
supercurrent, we must introduce \emph{moving pairons}, that is, pairons with finite
center-of-mass (CM) momenta. Creation operators for ``electron'' (1) and ``hole''
(2) pairons are defined by
\begin{equation}
B_{12}^{(1)\dagger} \equiv B_{\mathbf{k_1}\uparrow \mathbf{k_2}\downarrow}^{(1)\dagger}
\equiv c_1^\dagger c_2^\dagger, \quad
B_{34}^{(2)\dagger} = c_4^{(2)\dagger} c_3^{(2)\dagger}.
\end{equation}
We calculate the commutators among $B$ and $B^\dagger$, and obtain
\begin{equation}
\left[B_{12}^{(j)}, B_{34}^{(j)}\right] = 0,\qquad \left[B_{12}^{(j)}\right]^2 = 0,
\end{equation}
\begin{align}
&\left[B_{12}^{(j)}, B_{34}^{(j)\dagger}\right] \nonumber\\
&=
\begin{cases}
1 - n_1^{(j)} - n_2^{(j)} & \text{if } \mathbf{k_1} = \mathbf{k_3} \text{ and } \mathbf{k_2} = \mathbf{k_4}\\
c_2^{(j)}c_4^{(j)\dagger} & \text{if } \mathbf{k_1} = \mathbf{k_3} \text{ and } \mathbf{k_2} \neq \mathbf{k_4}\\
c_1^{(j)}c_3^{(j)\dagger} & \text{if } \mathbf{k_1} \neq \mathbf{k_3} \text{ and } \mathbf{k_2} = \mathbf{k_4}\\
0 & \text{otherwise}.
\end{cases}
\end{align}
Pairon operators of different types $j$ always commute:
\begin{equation}
\left[B^{(j)}, B^{(i)}\right] = 0\quad \text{if } j \neq i,
\end{equation}
and
\begin{equation}
n_1^{(j)} \equiv c_{\mathbf{k_1}\uparrow}^{(j)\dagger}
c_{\mathbf{k_1}\uparrow}^{(j)}, \qquad n_2^{(j)} \equiv
c_{\mathbf{k_2}\downarrow}^{(j)\dagger} c_{\mathbf{k_2}\downarrow}^{(j)}
\end{equation}
represent the number operators for ``electrons'' ($j = 1$) and ``holes'' ($j = 2$).

Let us now introduce the relative and net momenta ($\mathbf{k}, \mathbf{q}$) such that
\begin{equation}
\left.\begin{array}{l}
\mathbf{k} \equiv (\mathbf{k_1} - \mathbf{k_2}) / 2\\
\mathbf{q} \equiv \mathbf{k_1} + \mathbf{k_2}
\end{array}\right\}
\Longleftrightarrow
\left\{\begin{array}{l}
\mathbf{k_1} = \mathbf{k} + \mathbf{q} / 2\\
\mathbf{k_2} = -\mathbf{k} + \mathbf{q} / 2.
\end{array}\right.
\end{equation}
Alternatively we can represent pairon annihilation operators by
\begin{align}
B_{\mathbf{kq}}^{\prime(1)} &\equiv B_{k_1\uparrow k_2\downarrow}^{(1)}
\equiv c_{-\mathbf{k}+\mathbf{q}/2\downarrow}^{(1)}c_{\mathbf{k}+\mathbf{q}/2\uparrow}^{(1)}, \nonumber\\
B_{\mathbf{kq}}^{\prime(2)} &= c_{\mathbf{k}+\mathbf{q} / 2\uparrow}^{(2)}c_{-\mathbf{k}+\mathbf{q} / 2\downarrow}^{(2)}.
\end{align}
The prime on $B$ will be dropped hereafter. In the $\mathbf{k}$-$\mathbf{q}$ representation the
commutation relations are re-expressed as
\begin{equation}
\left[B_\mathbf{kq}^{(j)}, B_{\mathbf{k}'\mathbf{q}'}^{(i)}\right] = 0,\qquad \left[B_{\mathbf{kq}}^{(j)}\right]^2 = 0,
\label{eq:rar_as}
\end{equation}
\begin{align}
&\left[B_\mathbf{kq}^{(j)}, B_{\mathbf{k}'\mathbf{q}'}^{(i)\dagger}\right] \nonumber\\
&=
\begin{cases}
\left(1 - n_{\mathbf{k}+\mathbf{q}/2\uparrow} - n_{-\mathbf{k}+\mathbf{q}/2\downarrow}\right)\delta_{ji}\\
\text{if } \mathbf{k} = \mathbf{k}\boldsymbol{'} \text{ and } \mathbf{q} = \mathbf{q}\boldsymbol{'}.\\
c_{-\mathbf{k}+\mathbf{q}/2\downarrow}^{(j)} c_{-\mathbf{k}\boldsymbol{'}+\mathbf{q}\boldsymbol{'}/2\downarrow}^{(j)\dagger}\delta_{ji}\\
\text{if } \mathbf{k} + \mathbf{q}/2 = \mathbf{k}\boldsymbol{'}+\mathbf{q}\boldsymbol{'}/2\\
\text{and } -\mathbf{k} + \mathbf{q}/2 \neq -\mathbf{k}\boldsymbol{'}+\mathbf{q}\boldsymbol{'}/2.\\
c_{\mathbf{k}+\mathbf{q}/2\uparrow}^{(j)} c_{\mathbf{k}\boldsymbol{'}+\mathbf{q}\boldsymbol{'}/2\uparrow}^{(j)\dagger}\delta_{ji}\\
\text{if } \mathbf{k} +\mathbf{q}/2 \neq \mathbf{k}\boldsymbol{'}+\mathbf{q}\boldsymbol{'}/2\\
\text{and } -\mathbf{k} + \mathbf{q}/2 = -\mathbf{k}\boldsymbol{'}+\mathbf{q}\boldsymbol{'}/2.\\
0\\
\text{otherwise}.
\end{cases}
\label{eq:rar_as_2}
\end{align}
Using the new notations, we can write the full Hamiltonian as
\begin{align}
H = &\sum_{\mathbf{k},s}
\varepsilon_{\mathbf{k}}^{(1)}n_{\mathbf{k},s}^{(1)} + \sum_{\mathbf{k},s}
\varepsilon_{\mathbf{k}}^{(2)}n_{\mathbf{k},s}^{(2)} \nonumber\\
&- \sideset{}{'}{\sum}_\mathbf{k} \sideset{}{'}{\sum}_\mathbf{q} \sideset{}{'}{\sum}_{\mathbf{k}\boldsymbol{'}}v_0\Big[
B_{\mathbf{k}\mathbf{q}}^{(1)\dagger}B_{\mathbf{k}\boldsymbol{'}\mathbf{q}}^{(1)} +
B_{\mathbf{k}\mathbf{q}}^{(1)\dagger}B_{\mathbf{k}\boldsymbol{'}\mathbf{q}}^{(2)\dagger} + \nonumber\\
&B_{\mathbf{k}\mathbf{q}}^{(2)}B_{\mathbf{k}\boldsymbol{'}\mathbf{q}}^{(1)} +
B_{\mathbf{k}\mathbf{q}}^{(2)}B_{\mathbf{k}\boldsymbol{'}\mathbf{q} }^{(2)\dagger}\Big].
\label{eq:full_Hamiltonian}
\end{align}
This is the full Hamiltonian for the system, which can describe moving pairons
as well as stationary pairons. Here, the prime on the summations indicates the
restriction arising from the phonon exchange attraction, see below. The connection
with BCS Hamiltonian \cite{Bardeen} will be discussed in Sec.~\ref{se:BEC_Pairons}.

\section{MOVING PAIRONS}\label{se:Moving_Pairons}

The phonon exchange attraction is in action for any pair of electrons
near the Fermi surface. In general the bound pair has a net momentum,
and hence it moves. The energy $w_q$ of a moving pairon can be obtained
from:
\begin{align}
w_q a(\mathbf{k}, \mathbf{q}) = &[\varepsilon(|\mathbf{k}+\mathbf{q}/2|) +
\varepsilon(|-\mathbf{k}+\mathbf{q}/2|)] a(\mathbf{k}, \mathbf{q}) \nonumber\\
&- \frac{v_0}{(2\pi\hbar)^2} \int' d^2k' a(\mathbf{k}\boldsymbol{'}, \mathbf{q}),
\label{eq:moving_pairon_energy}
\end{align}
which is Cooper's equation, Eq.~(1) of his 1965 Physical Review
Letter.\ \cite{Cooper} The prime on the $k'$-integral means the
restriction on the integration domain arising from the phonon exchange
attraction, see below. We note that the net momentum $\mathbf{q}$ is a
constant of motion, which arises from the fact that the phonon exchange
attraction is an internal force, and hence cannot change the net
momentum. The \emph{pair wavefunctions} $a(\mathbf{k, q})$ are coupled
with respect to the other variable $\mathbf{k}$, meaning that the exact
(or energy-eigenstate) pairon wavefunctions are superpositions of the
pair wavefunctions $a(\mathbf{k, q})$.

Equation \eqref{eq:moving_pairon_energy} can be solved as follows. We assume that the energy $w_q$ is
negative:
\begin{equation}
w_q < 0.
\end{equation}
Then,
\begin{equation}
\varepsilon(|\mathbf{k} + \mathbf{q}/2|) + \varepsilon(| -\mathbf{k} + \mathbf{q}/2|) - w_q > 0.
\end{equation}
Rearranging the terms in Eq.~\eqref{eq:moving_pairon_energy} and dividing by $\varepsilon(|\mathbf{k}+\mathbf{q}/2|)+\varepsilon(|-\mathbf{k}+\mathbf{q}/2|)-w_q$,
we obtain
\begin{equation}
a(\mathbf{k}, \mathbf{q}) = [\varepsilon(|\mathbf{k} + \mathbf{q}/2|) + \varepsilon(| - \mathbf{k} + \mathbf{q}/2|) - w_q]^{-1}C(\mathbf{q}),
\label{eq:Ead_by}
\end{equation}
where
\begin{equation}
C(\mathbf{q}) \equiv
\frac{v_0}{(2\pi\hbar)^2} \int'd^2k' a(\mathbf{k}\boldsymbol{'}, \mathbf{q}),
\end{equation}
which is $k$-independent.

Introducing Eq.~\eqref{eq:Ead_by} in Eq.~\eqref{eq:moving_pairon_energy}, and dropping the common factor $C(\mathbf{q})$, we
obtain
\begin{align}
1 = &\frac{v_0}{(2\pi\hbar)^2} \int'd^3k [\varepsilon(|\mathbf{k} + \mathbf{q}/2|) + \varepsilon(| - \mathbf{k} + \mathbf{q}/2|) + \nonumber\\
&|w_q|]^{-1}.
\label{eq:dtc_factor}
\end{align}

\begin{figure}
\centering
\includegraphics{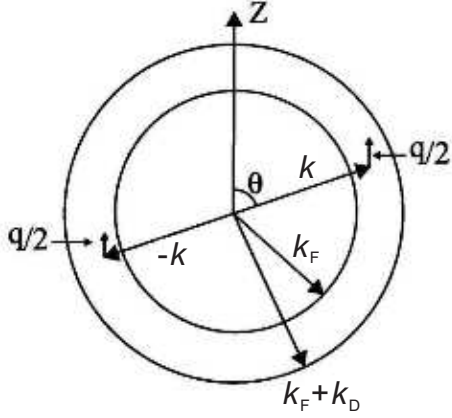}
\caption{The range of the integration variables $(k,\theta)$ is limited to a spherical shell of thickness $k_\text{D}$.}\label{fg:Figure4}
\end{figure}

We now assume a free electron moving in 3D. The Fermi surface is a sphere of
the radius (momentum)
\begin{equation}
k_\text{F} \equiv (2m_1\varepsilon_\text{F})^{1/2},
\end{equation}
where $m_1$ represents the effective mass of an electron. The energy $\varepsilon(|\mathbf{k}|)$ is given by
\begin{equation}
\varepsilon(|\mathbf{k}|) \equiv \varepsilon_k = \frac{k^2 - k_\text{F}^2}{2m_1}.
\end{equation}
The prime on the $k$-integral in Eq.~\eqref{eq:dtc_factor} means the restriction:
\begin{equation}
0 < \varepsilon(|\mathbf{k} + \mathbf{q}/2|),\quad \varepsilon(|-\mathbf{k} + \mathbf{q}/2|) < \hbar\omega_\text{D}.
\end{equation}
We may choose the polar axis along q as shown in Fig.~\ref{fg:Figure4}.%
The integration with respect to the azimuthal angle simply yields the
factor $2\pi$. The $k$-integral can then be expressed by
\begin{align}
\frac{(2\pi\hbar)^3}{v_0} &= 4\pi \int_0^{\pi/2}
d\theta \sin\theta \int_{k_\text{F}+\frac{1}{2}q \cos\theta}^{k_\text{F}+k_\text{D}-\frac{1}{2}q \cos\theta}\cdot \nonumber\\
&\frac{k^2dk}{|w_q| + 2\varepsilon_k + (4m_1)^{-1}q^2},
\end{align}
\begin{equation}
k_\text{D} \equiv m_1\hbar\omega_\text{D}k_\text{F}^{-1}.
\end{equation}
After performing the integration and taking the small-$q$ and small-$(k_\text{D}/k_\text{F})$ limits,
we obtain
\begin{equation}
w_q = w_0 + \frac{v_\text{F}}{2}q,
\label{eq:sas_limits}
\end{equation}
where the pairon ground-state energy $w_0$ is given by
\begin{equation}
w_0 = \frac{-2\hbar\omega_\text{D}}{\exp[2/v_0\mathcal{N}(0)] - 1}.
\label{eq:wig_by}
\end{equation}
As expected, the zero-momentum pairon has the lowest energy $w_0$. The
excitation energy is continuous with no energy gap. Equation
\eqref{eq:sas_limits} was first obtained by Cooper (unpublished), and it
is recorded in Schrieffer's book \cite{Schrieffer}, Eq. (2.15). The
energy $w_q$ increases \emph{linearly} with momentum (magnitude) $q$ for
small $q$. This behavior arises from the fact that the density of states
is strongly reduced with the increasing momentum $q$ and dominates the
$q^2$ increase of the kinetic energy. The linear dispersion relation
means that a \emph{pairon moves likes a massless particle} with a common
speed $v_\text{F}/2$. This relation plays a vital role in the B-E
condensation of pairons (see next section).

Such a linear energy-momentum relation is valid for
pairons moving in any dimension (D). However, the coefficients slightly
depend on the dimensions; in fact
\begin{equation}
w_q = w_0 + cq.
\label{eq:tdi_fact}
\end{equation}
$c/v_\text{F} = 1/2$ and $2/\pi$ for 3D and 2D, respectively.

\section{THE BOSE-EINSTEIN CONDENSATION OF PAIRONS}\label{se:BEC_Pairons}

\begin{figure*}
\centering
\includegraphics{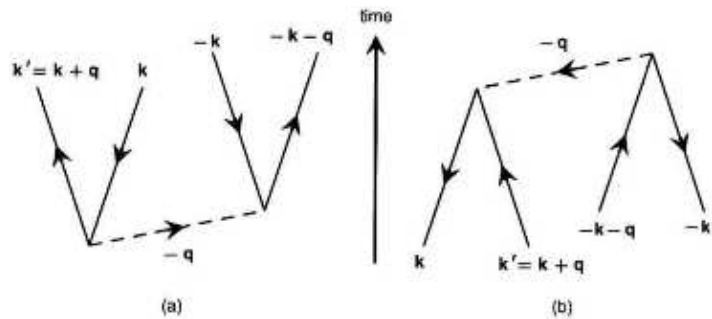}
\caption{Feynman diagrams representing (a) pair-creation of $\pm$ ground pairons from the physical
vacuum, and (b) pair annihilation. The time is measured upwards.}\label{fg:Figure5}
\end{figure*}

In Sec.~\ref{se:Full_Hamiltonian}, we saw that the pair operators ($B,B^\dagger$) appearing in the full Hamiltonian
$H$ in Eq.~\eqref{eq:full_Hamiltonian} satisfy rather complicated commutator relations in Eqs.\ \eqref{eq:rar_as} and
\eqref{eq:rar_as_2}. In particular part of Eq.~\eqref{eq:rar_as}
\begin{equation}
\left[B_{\mathbf{k}0}^\dagger\right]^2 \equiv \left[b_\mathbf{k}^\dagger\right]^2 =
\left[c_{-\mathbf{k}\uparrow}^\dagger c_{\mathbf{k}\downarrow}^\dagger\right]^2 = 0
\end{equation}
reflect the fermionic natures of the constituting electrons. Here,
$B_{\mathbf{k}0}^\dagger \equiv b_\mathbf{k}^\dagger$ represents
creation operator for zero momentum pairons. BCS~\cite{Bardeen} studied the ground-state of a
superconductor, starting with the reduced Hamiltonian $H_0$, which is
obtained from the Hamiltonian $H$ in Eq.~\eqref{eq:full_Hamiltonian} by
retaining the zero momentum pairons with $q = 0$, written in terms of
$b$ by letting $B_{\mathbf{k}0}^{(j)} = b_\mathbf{k}^{(j)}$,
\begin{align}
H_0 = &\sum_\mathbf{k} 2\varepsilon_\mathbf{k}^{(1)} b_\mathbf{k}^{(1)\dagger}b_\mathbf{k}^{(1)} +
\sum_\mathbf{k} 2\varepsilon_\mathbf{k}^{(2)} b_\mathbf{k}^{(2)\dagger}b_\mathbf{k}^{(2)} \nonumber\\
&- \sideset{}{'}\sum_\mathbf{k} \sideset{}{'}\sum_\mathbf{k'}v_0 \Big[
b_\mathbf{k'}^{(1)\dagger}
b_\mathbf{k}^{(1)} +
b_\mathbf{k'}^{(1)\dagger}
b_\mathbf{k}^{(2)\dagger} + \nonumber\\
&b_\mathbf{k'}^{(2)}
b_\mathbf{k}^{(1)} +
b_\mathbf{k'}^{(2)}
b_\mathbf{k}^{(2)\dagger}\Big].
\end{align}
Here, we expressed the ``electron'' and ``hole'' kinetic energies in
terms of pairon operators. The reduced Hamiltonian $H_0$ is bilinear in
pairon operators $(b, b^\dagger)$, and can be diagonalized exactly. BCS
obtained the ground-state energy $E_0$ as
\begin{equation}
E_0 = \hbar\omega_\text{D}\mathcal{N}(0)w_0,
\label{eq:ground_state_energy}
\end{equation}
where $\mathcal{N}(0)$ is the density of states at the Fermi energy. The $w_0$ is the ground-state
energy of the pairon, see Eq.~\eqref{eq:wig_by}. Equation \eqref{eq:ground_state_energy} means simply that the ground
state energy equals the numbers of pairons times the ground-state energy $w_0$
of the pairon. Our Hamiltonian $H$ in Eq.~\eqref{eq:full_Hamiltonian} is reduced to the original BCS
Hamiltonian (see Ref.~\cite{Bardeen}, Eq. (24)). There is an important difference in the
definition of ``electron'' and ``hole'' here. BCS called the quasi-electron whose energy
is higher (lower) than the Fermi energy $\varepsilon_\text{F}$, the ``electron'' (``hole''). In our theory
the ``electrons'' (``holes'') are defined as quasiparticles generated above (below) the
Fermi energy \emph{and} circulates counterclockwise (clockwise) viewed from the tip of
an external magnetic field vector $\mathbf{B}$. They are generated, depending on the energy
contour curvature signs. For example, only ``electrons'' (``holes'') are generated for
a circular Fermi surface with negative (positive) curvature whose inside (outside) is
filled with electrons. Since the phonon has no charge, the phonon exchange cannot
change the net charge. The pairing interaction terms in
Eq.~\eqref{eq:full_Hamiltonian} conserve the charge.
The term $-v_0 B_{\mathbf{kq}s}^{(1)\dagger} B_{\mathbf{k}\boldsymbol{'}\mathbf{q}s}^{(1)} $,
where $v_0 \equiv \left|V_\mathbf{q} V_\mathbf{q}'\right|(\hbar\omega_0 A)^{-1}$,
$A =$ sample area, generates a transition in the ``electron'' states.
Similarly, the exchange of a phonon generates a transition in the
``hole'' states represented by $-v_0 B_{\mathbf{kq}s}^{(2)\dagger}
B_{\mathbf{k}\boldsymbol{'}\mathbf{q}s}^{(2)\dagger}$. The phonon exchange can also
pair-create or pair-annihilate ``electron'' (``hole'') pairons, and the
effects of these processes are represented by
$-v_0 B_{\mathbf{kq}s}^{(1)\dagger}B_{\mathbf{k}\boldsymbol{'}{q}s}^{(2)\dagger}$,
$-v_0
B_{\mathbf{kq}s}^{(1)}B_{\mathbf{k}\boldsymbol{'}\mathbf{q}s}^{(2)}$, as
shown in Feynman diagrams in Figs.~\ref{fg:Figure5}(a) and
\ref{fg:Figure5}(b).%
At $0$\,K the system must have equal numbers of $-$ ($+$) zero-momentum
(ground) pairons.

To describe a supercurrent, we must introduce moving pairons. We now show
that the center-of-masses of the pairons move as bosons. That is, the
number operator of pairons having net momentum $\mathbf{q}$
\begin{equation}
n_\mathbf{q} \equiv \sum_\mathbf{k}
n_\mathbf{kq} = \sum_\mathbf{k} B_\mathbf{kq}^\dagger B_\mathbf{kq}
\end{equation}
have the eigenvalues
\begin{equation}
n_\mathbf{q}' = 0,1,2,\dots.
\label{eq:eigenvalues}
\end{equation}
The number operator for the pairons in the state $(\mathbf{k, q})$ is
\begin{equation}
n_\mathbf{kq} \equiv B_\mathbf{kq}^\dagger B_\mathbf{kq} = c_{\mathbf{k}+\mathbf{q}/2}^\dagger
c_{-\mathbf{k}+\mathbf{q}/2}^\dagger c_{-\mathbf{k}+\mathbf{q}/2}c_{\mathbf{k}+\mathbf{q}/2},
\end{equation}
where we omitted the spin indices. Its eigenvalues are limited to zero or one:
\begin{equation}
n_\mathbf{kq}' = 0 \text{ or } 1.
\label{eq:eigenvalues_zero_one}
\end{equation}
To explicitly see this property in Eq.~\eqref{eq:eigenvalues}, we introduce
\begin{equation}
B_\mathbf{q} \equiv \sum_\mathbf{k} B_\mathbf{kq}
\end{equation}
and obtain
\begin{align}
[B_\mathbf{q}, n_\mathbf{q}] &= \sum_\mathbf{k}
\left(1 - n_{\mathbf{k}+\frac{1}{2}\mathbf{q}} - n_{-\mathbf{k}+\frac{1}{2}\mathbf{q}}\right) B_\mathbf{kq} = B_\mathbf{q}, \nonumber\\
\left[n_\mathbf{q}, B_\mathbf{q}^\dagger\right] &= B_\mathbf{q}^\dagger.
\label{eq:and_obtain}
\end{align}
Although the occupation number $n_\mathbf{q}$ is not connected with $B_\mathbf{q}$
as $n_\mathbf{q} \neq B_\mathbf{q}^\dagger B_\mathbf{q}$, the
eigenvalues $n_\mathbf{q}'$ of $n_\mathbf{q}$ satisfying
Eq.~\eqref{eq:and_obtain} can be shown straightforwardly to yield \cite{Dirac}
\begin{equation}
n_\mathbf{q}' = 0,1,2, \dots,
\end{equation}
with the eigenstates
\begin{equation}
|0\rangle, \quad |1\rangle = B_\mathbf{q}^\dagger
|0\rangle, \quad |2\rangle = B_\mathbf{q}^\dagger
B_\mathbf{q}^\dagger|0\rangle, \dots.
\end{equation}

In summary, pairons with both $\mathbf{k}$ and $\mathbf{q}$ specified
are subject to the Pauli exclusion principle, see Eq.~\eqref{eq:eigenvalues_zero_one}.
Yet, the occupation numbers $n_\mathbf{q}'$ of pairons having a CM momentum
$\mathbf{q}$ are $0,1,2, \dots$.

The most important signature of many bosons is the \emph{Bose-Einstein
Condensation} (BEC). Earlier we showed that the pairon moves in 2D with the
linear dispersion relation, see \eqref{eq:sas_limits}:
\begin{equation}
w_p = w_0 + (2/\pi)v_\text{F}p \equiv w_0 + c p,
\end{equation}
where we designated the pairon net momentum (magnitude) by the more familiar
$p$ rather than $q$.

Let us consider a 2D system of free bosons having a linear dispersion relation:
$\varepsilon = cp$, $c \equiv (2/\pi)v_\text{F}$. The number of bosons, $N$,
and the Bose distribution function
\begin{equation}
f_\text{B}(\varepsilon; \beta, \mu) \equiv \frac{1}{e^{\beta(\varepsilon - \mu)} - 1} \equiv 
f_\text{B}(\varepsilon)\ (> 0)
\end{equation}
are related by
\begin{equation}
N = \sum_p f_\text{B}(\varepsilon_p; \beta, \mu) = N_0 + \sideset{}{'}{\sum}_{\substack{p\\ \varepsilon_p > 0}}
f_\text{B}(\varepsilon_p),
\label{eq:ar_by}
\end{equation}
where $\mu$ is the chemical potential, $\beta \equiv (k_\text{B}T)^{-1}$, and
\begin{equation}
N_0 \equiv (e^{-\beta\mu} - 1)^{-1}
\end{equation}
is the number of zero-momentum bosons. The prime on the summation in Eq.~\eqref{eq:ar_by}
indicates the omission of the zero-momentum state. For notational
convenience, we write
\begin{equation}
\varepsilon = cp = (2/\pi)v_\text{F}p\ (> 0).
\label{eq:ncw_w}
\end{equation}

We divide Eq.~\eqref{eq:ar_by} by the normalization area $L^2$, and take the \emph{bulk limit}:
\begin{equation}
N \rightarrow \infty,\ L \rightarrow \infty\quad \text{while } N L^{-2} \equiv n.
\end{equation}
We then obtain
\begin{equation}
n - n_0 \equiv \frac{1}{(2\pi\hbar)^2} \int d^2p f_\text{B}(\varepsilon),
\label{eq:Wt_obtain}
\end{equation}
where $n_0 \equiv N_0/L^2$ is the number density of zero-momentum bosons and $n$ the total
boson density. After performing the angular integration and changing integration
variables, we obtain from Eq.~\eqref{eq:Wt_obtain}:
\begin{align}
2\pi\hbar^2 c^2\beta^2 (n - n_0) &= \int_0^\infty dx\frac{x}{\lambda^{-1}e^x - 1} \nonumber\\
&(x \equiv \beta\varepsilon),
\label{eq:wof_Eq}
\end{align}
where the fugacity
\begin{equation}
\lambda \equiv e^{\beta\mu}\ (< 1).
\end{equation}
is less than unity for all temperatures. After
expanding the integrand in Eq.~\eqref{eq:wof_Eq} in powers of $\lambda e^{-x}\ (< 1)$,
and carrying out the $x$-integration, we obtain
\begin{equation}
n_x \equiv n - n_0 = \frac{k_\text{B}^2T^2 \phi_2(\lambda)}{2\pi\hbar^2c^2},
\label{eq:cot_x}
\end{equation}
\begin{equation}
\phi_2(\lambda) \equiv \sum_{k=1}^\infty \frac{\lambda^k}{k^2}\qquad (0 \leq \lambda \leq 1).
\end{equation}
Equation \eqref{eq:cot_x} gives a relation among $\lambda$, $n$, and $T$.

The function $\phi_2(\lambda)$ monotonically increases from zero to the
maximum value
\begin{equation}
\phi_2(1) = 1.645
\end{equation}
as $\lambda$ is raised from zero to one. In the
low-temperature limit, $\lambda = 1$, $\phi_2(\lambda) = \phi_2(1) =
1.645$, and the density of excited bosons, $n_x$, varies as $T^2$ as
seen from Eq.~\eqref{eq:cot_x}. This temperature behavior of $n_x$ persists as long
as the right-hand-side (r.h.s.) of Eq.~\eqref{eq:cot_x} is smaller than $n$; the
\emph{critical temperature} $T_c$ occurs at $n_c = k_\text{B}^2 T_c^2 \phi_2(1) /
2\pi\hbar^2c^2$. Solving this, we obtain
\begin{equation}
k_\text{B}T_c = 1.954 \hbar c n^{1/2}\ \left(=1.24 \hbar v_\text{F} n^{1/2}\right).
\label{eq:anS_this}
\end{equation}

The BEC of pairons moving in 2D occurs at a finite temperature. This appears
to contradict with Hohenberg's theorem (no long range order in 2D). But this
theorem is proved under the assumption of the f-sum rule arising from the mass
conservation. The pairons move massless with the linear dispersion relation [see
Eq.\ \eqref{eq:ncw_w}], and hence they are not subject to Hohenberg's
theorem.~\cite{Hohenberg}

If the temperature is raised beyond $T_c$, the density of zero momentum bosons,
$n_0$, becomes vanishingly small, and the fugacity $\lambda$ can be determined from
\begin{equation}
n = \frac{k_\text{B}T^2 \phi_2(\lambda)}{2\pi\hbar^2c^2},\quad T > T_c.
\label{eq:cbd_from}
\end{equation}

In summary, the fugacity $\lambda$ is equal to unity in the condensed region: $T < T_c$,
and it becomes smaller than unity for $T > T_c$, where its value is determined from
Eq.~\eqref{eq:cbd_from}.

Formula \eqref{eq:anS_this} for the critical temperature $T_c$ is distinct from the famous BCS
formula
\begin{equation}
3.53\,k_\text{B}T_c = 2\triangle_0,
\label{eq:BCS_formula}
\end{equation}
where $\triangle_0$ is the zero temperature electron energy gap in the
weak coupling limit. The electron energy gap $\triangle(T)$ and the
pairon ground-state energy $w_0$ both depend on the phonon-exchange
coupling energy parameter $v_0$, which appears in the starting
Hamiltonian $H$ in Eq.~\eqref{eq:full_Hamiltonian}. The energy $w_0$ is
negative (bound-state energy). Hence, this $w_0$ cannot be obtained by
the perturbation theory. The connection between $w_0$ and $v_0$ is very
complicated. This makes it difficult to discuss the critical temperature
$T_c$ based on the BCS relation \eqref{eq:BCS_formula}. Unlike the BCS
formula, formula \eqref{eq:anS_this} is directly connected with the
measurable quantities: the pairon density $n_0$ and the Fermi speed
$v_\text{F}$.

We emphasize here that both formulas \eqref{eq:anS_this} and
\eqref{eq:BCS_formula} were derived, starting with the Hamiltonian $H$
in Eq.~\eqref{eq:full_Hamiltonian} and following statistical mechanical
calculations, see the reference \cite{Fujita10} for details.

\section{ZERO BIAS ANOMALY}\label{se:ZBA}

The unusual current-dips at zero bias in Fig.~\ref{fg:Figure1} is often called the \emph{zero bias
  anomaly} (ZBA). This effect is clearly seen in (a) low resistance
contacts LRC sample. The differential conductance $dI/dV$ increases
with increasing bias, reaching a maximum at $V \sim 100$\,mV. With a
further bias increase, $dI/dV$ drops dramatically. See (a), the upper
panel in Fig.~\ref{fg:Figure1}. We will show that the ZBA arises from
the break-down of the superconducting state of the system.

With no bias, the nanotube's wall below $\sim 100$\,K is in a
superconducting state. If a small bias is applied, then the system is
charged, positively or negatively depending on the polarity of the
external bias. The applied bias field will not affect the neutral
supercurrent but can accelerate the charges at the outer side of the
carbon wall. The resulting normal currents carried by conduction
electrons are scattered by impurities and phonons. The phonon population
changes with temperature, and hence the phonon scattering is
temperature-dependent. The normal electric currents along the tube
length generate circulating magnetic fields, which destroy eventually
the supercurrent running in the wall at a high enough bias. Thus, the
current $I$ ($\mu A$) versus the voltage $V$ (mV) is non-linear near the
origin because of the supercurrents running in the wall. The
differential conductance $dI/dV$ is very small and nearly constant
(superconducting) for $V < 10$\,mV in the HRC sample, see the lower
panel in Fig.~\ref{fg:Figure1}. We stress that if the ballistic electron
model~\cite{Saito} is adopted, then the scatterings by phonons cannot be
discussed. The non-linear $IV$ curves below $150$\,K mean that the
carbon wall is superconducting. Thus, the clearly visible temperature
effects for both LRC and HRC samples arise from the phonon scattering.
We assumed that the system is superconducting below $\sim 150$\,K. The
ZBA arises only from the superconducting state. The superconducting
critical temperature $T_c$ must then be higher than $150$\,K. An
experimental check of $T_c$ is highly desirable.


\begin{thebibliography}{99}

\bibitem{Yao}

Z.~Yao, C.L.~Kane, and C.~Dekker, \emph{Phys.\ Rev.\ Lett.}\ \textbf{84}, 2941 (2000).

\bibitem{Saito}

R.~Saito, G.~Dresselhaus, and M.S.~Dresselhaus, \emph{Physical Properties of
Carbon Nanotubes} (Imperial College Press, London, 1998), pp.~35--39,
pp.~155--156, pp.~139--144.

\bibitem{Iijima1}

S.~Iijima, \emph{Nature}, \textbf{354}, 56 (1991).

\bibitem{Wigner}

E.~Wigner and F.~Seitz, \emph{Phys.\ Rev.}\ \textbf{43}, 804 (1933).

\bibitem{Iijima2}

S.~Iijima and T.~Ishibashi, \emph{Nature}, \textbf{363}, 603 (1993);
D.S.~Bethune, \emph{et~al}., \emph{Nature}, \textbf{363}, 605 (1993).

\bibitem{Fujita1}

S.~Fujita, Y.~Takato, and A.~Suzuki, \emph{Mod.\ Phys.\ Lett.}\ \emph{25}, 223 (2011).

\bibitem{Fujita2}

S.~Fujita, Y.~Takato, S.~Godoy, and A.~Suzuki, arXiv:1003.5231v1
[cond-mat.mes-hall] (2010).

\bibitem{Moriyama}

S.~Moriyama, K.~Toratani, D.~Tsuya, M.~Suzuki, Y.~Aoyagi, and
K.~Ishibashi, \emph{Physica E}, \textbf{24}, 46 (2004).

\bibitem{Tans1}

S.J.~Tans, M.H.~Devoret, H.~Dai, A.~Thess, R.E.~Smalley, L.J.~Geerligs,
and C.~Dekker, \emph{Nature}, \textbf{386}, 474 (1997).

\bibitem{Tans2}

S.J.~Tans, A.R.M.~Verschueren, and C.~Dekker, \emph{Nature},
\textbf{393}, 49 (1998).

\bibitem{Cooper}

L.N.~Cooper, \emph{Phys.\ Rev.}\ \textbf{104}, 1189 (1956).

\bibitem{Bachtold}

A.~Bachtold, M.S.~Fuhrer, S.~Plyasunov, M.~Forero, Erik H.~Anderson,
A.~Zettl, and Paul L.~McEuen, \emph{Phys.\ Rev.\ Lett.}\ \textbf{84}, 6082 (2000).

\bibitem{Ashcroft}

N.W.~Ashcroft and N.D.~Mermin, \emph{Solid State Physics} (Saunders,
Philadelphia, 1976), pp.~228--229.

\bibitem{Harrison}

W.A.~Harrison, \emph{Solid State Theory} (Dover, New York, 1980), pp.~390--393.

\bibitem{Frohlich}

H.~Fr\"ohlich, \emph{Phys.\ Rev.}\ \textbf{79}, 845 (1950); Proc

\bibitem{Bardeen}

J.~Bardeen, L.N.~Cooper, and J.R.~Schrieffer, \emph{Phys.\ Rev.}\ \textbf{108}, 1175 (1957).

\bibitem{Fujita5}

S.~Fujita, \emph{J.\ Supercond.}\ \textbf{4} (1991) 297.

\bibitem{Fujita6}

S.~Fujita, \emph{J.\ Supercond.}\ \textbf{5} (1992) 83.

\bibitem{Fujita7}

S.~Fujita and S.~Watanabe, \emph{J.\ Supercond.}\ \textbf{5} (1992) 219.

\bibitem{Fujita8}

S.~Fujita and S.~Watanabe, \emph{J.\ Supercond.}\ \textbf{6} (1993) 75.

\bibitem{Fujita9}

S.~Fujita and S.~Godoy, \emph{J.\ Supercond.}\ \textbf{6} (1993) 373.

\bibitem{Schrieffer}

J.R.~Schrieffer, \emph{Theory of Superconductivity} (Benjamin, New York, 1964), p.~33.

\bibitem{Dirac}

P.A.M.~Dirac, \emph{Principle of Quantum Mechanics}, 4th edn. (Oxford University Press,
London, 1958), pp.~211, 136–-138, 37, 253–-257.

\bibitem{Hohenberg}

P.C.~Hohenberg, \emph{Phys.\ Rev.}\ \textbf{158} (1967).

\bibitem{Mermin}

N.D.~Mermin and H.~Wagner, \emph{Phys.\ Rev.\ Lett.}\ \textbf{17} (1966) 1133.

\bibitem{Fujita10}

S.~Fujita, K.~Ito, and S.~Godoy, \emph{Quantum Theory of Conducting
Matter: Superconductivity} (Springer, New York, 2009), pp.~79--81.

\end{thebibliography}
\end{document}